%% file: rv2014final.tex
\documentclass{llncs}
\usepackage{llncsdoc}
\usepackage{amsfonts}
\usepackage{amsmath}
\usepackage{amstext}
\usepackage{amssymb}
\usepackage{latexsym}
\usepackage[ruled]{algorithm}
\usepackage{algpseudocode}
\usepackage{mathtools}
\usepackage{todonotes}
\usepackage{txfonts}
\usepackage{fainekos-macros}
\usepackage{booktabs} 
\usepackage{bigstrut} 
\usepackage{graphicx}
\usepackage{multicol}
\usepackage{multirow}

\usepackage{xcolor,colortbl}
\usepackage{vwcol}  
\graphicspath{{figures/}}

\usepackage{color,soul}

\usepackage{ifthen}
\newboolean{HIGHCOMM}
\setboolean{HIGHCOMM}{false}
\newcommand \responsehighlight[1]{\ifthenelse{\boolean{HIGHCOMM}}{\textcolor{red}{#1}}{#1}}
\newboolean{HIGHLIGHT}
\setboolean{HIGHLIGHT}{false}
\newcommand{\yhl}[1]{\ifthenelse{\boolean{HIGHLIGHT}}{\hl{#1}}{#1}}

\usepackage{stmaryrd}

\def\dle{[\![} \def\dri{]\!]} 
\def\Un{\, \Uc} 

\renewcommand{\Re}{\mathbb{R}}

\begin{document}

\title{On-Line Monitoring for Temporal Logic Robustness}

\author{Adel Dokhanchi \and Bardh Hoxha \and Georgios Fainekos} 

\institute{School of Computing, Informatics and Decision Systems Engineering\\ Arizona State University\\
\email{\{adokhanc,bhoxha,fainekos\}@asu.edu} }

\maketitle             

\begin{abstract} 
In this paper, we provide a Dynamic Programming algorithm for on-line monitoring of the state robustness of Metric Temporal Logic specifications with past time operators. We compute the robustness of MTL with unbounded past and bounded future temporal operators (\bMITL) over sampled traces of Cyber-Physical Systems. We implemented our tool in Matlab as a Simulink block that can be used in any Simulink model. We experimentally demonstrate that the overhead of the \bMITL robustness monitoring is acceptable for certain classes of practical specifications.
\end{abstract}

\input{intro}

\input{problem}

\input{prelim}

\input{finitehorizon}

\input{experiments}

\input{casestudy}

\input{conclusions}

\paragraph*{Acknowledgments}
This work was partially supported by NSF awards CNS 1116136 and CNS 1319560. 
The authors would also like to thank the anonymous reviewers for the very detailed comments.

\bibliographystyle{splncs}
\bibliography{fatesrv}
\newpage
\input{appendix}

\end{document}

%% file: intro.tex
\section{Introduction}

Modern airplanes, automobiles and medical devices are prime examples of safety critical Cyber-Physical Systems (CPS).
Nowadays, the majority of safety critical functions in such systems is controlled by embedded computers.
Due to the critical nature of these components, it is of paramount importance to verify the functional correctness of the embedded software.
However, as the number of computer controlled components increases so does the complexity of the verification of functional correctness.
Moreover, the verification problem of most classes of CPS is even an undecidable problem \cite{hs_gf:Alur95algorithmic}.

As an alternative to verification and off-line testing, runtime monitoring has been proposed.
The underlying idea is that given a set of formal requirements, these requirements are analyzed at runtime by an independent monitor and if a violation is detected, it is reported to a supervisor.
The supervisor can then decide on remedial actions to fix the problem or reduce its impact to the system.
The monitoring problem has been extensively studied \cite{FinkbeinerK09,HavelundR01ase,Havelund02,HavelundR04,KristoffersenPA03,MalerNickovic04,ReinbacherRS14tacas,Rosu01,TanKSL04,Thati05entcs,BasinKZ11rv,geilen01construction,MalerNP06} for the cases where the formal requirements are expressed in Linear Temporal Logic (LTL) \cite{Pnueli77sfcs} or in Metric Temporal Logic (MTL) \cite{Koymans90}.

In this paper, we revisit the MTL runtime monitoring problem when targeted to CPS.
In particular, we claim that the classical Boolean semantics (or even three valued semantics) are not sufficiently informative for CPS behaviors.
For instance, consider the specification ``{\it After a takeoff command is received, then reach altitude of 600ft within 5 minutes}" for an autonomous Unmanned Aerial Vehicle (UAV) as introduced in \cite{ReinbacherRS14tacas}.
Clearly, knowing that the specification failed or passed at runtime is important.
However, more useful information from the perspective of the supervisor would be the knowledge of how far is the aircraft from satisfying the requirement.
More specifically, -10ft from the requirement of 600ft at 1 min away from the 5 min threshold should potentially be less alarming than -100ft at exactly the same time. 
A supervisor that has a model of the dynamics of the aircraft can determine whether the UAV can climb 100ft within 1 min or not.
We remark that the determination of the climb rate can only occur at runtime since this depends on the atmospheric parameters, the payload of the UAV, etc. Hence, the climb rate cannot be a precomputed parameter unless it is very conservatively set.

Our goal is to construct MTL monitors for estimating the robustness of satisfaction \cite{FainekosP06fates,FainekosP09tcs,DonzeAF2013stl}.
Temporal logic robustness gives a quantitative interpretation of satisfaction of an MTL formula.
In detail, if an MTL formula valuates to positive robustness $\varepsilon$, then the specification is true and, moreover, the state sequences can tolerate perturbations up to $\varepsilon$ and still satisfy the specification.
Similarly, if the robustness is negative, then the specification is false and, moreover, the state sequences under $\varepsilon$ perturbations still do not satisfy the specification.
Thus, robust semantics can be used to give quantitative values to the satisfaction of MTL formulas when the target is CPS.

The challenge here is that automata based monitors \cite{geilen01construction,MalerNP06} cannot be synthesized for computing the robustness valuations.
Therefore, formula rewriting methods \cite{Thati05entcs} or dynamic programming \cite{Rosu01} methods must be used.
Here, we take the latter approach for combined unbounded past time and bounded future time MTL specifications.
Since we are working with CPS, we assume that it is possible - if desired - to have a model predictive component in the system \cite{GarciaPM89automatica} which will provide a finite horizon prediction of the system behavior. 
That finite horizon prediction could be appended with the observed system behavior to provide a robustness estimate of a likely system behavior.
Hence, it becomes possible to monitor specifications such as ``{\it \responsehighlight{If at anytime in the past a takeoff command is issued, then within 5 min the altitude of 600ft is reached}}".
Thus, such requirements can now be monitored using only the actual observed system behavior or the observed system behavior with the predicted system behavior.

\responsehighlight{Our contributions in this paper are as follows: We provide a dynamic programming algorithm for on-line monitoring of the robustness metric of MTL formulas with bounded future and unbounded past.} In addition,
we provide a Matlab/Simulink \responsehighlight{toolbox} that can be used in any Simulink model for runtime monitoring of MTL robustness. \responsehighlight{The memory usage of our method is bounded and its runtime overhead is negligible for practical applications.}
Additional benefits in utilizing an on-line monitor are that it can be used in temporal logic testing algorithms \cite{Abbas13tecs,JinEtAl13hscc}, where it may be desirable that the simulation stops as soon as the property is \responsehighlight{violated, as well as in feedback control for MTL specifications. Although temporal logic robustness has been considered in previous works \cite{FainekosP06fates,FainekosP09tcs,DonzeAF2013stl}, the solutions were provided for off-line testing. To the best of our knowledge, this is the first attempt to solve the on-line MTL robustness monitoring problem efficiently.}

%% file: problem.tex
\section{Problem Formulation}

\begin{figure}[t]
\begin{center}
\includegraphics[width=12cm]{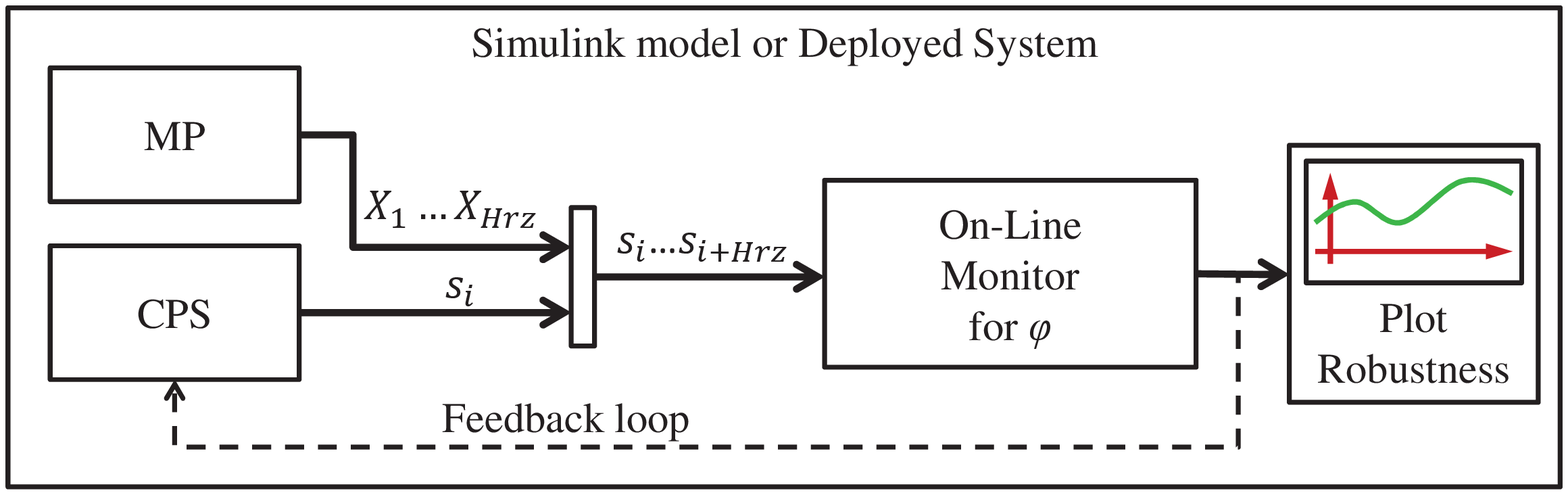} 
\end{center}
\vspace{-15pt}
\caption{Overview of the solution of the \bMITL on-line monitoring problem. The monitored robustness values could be used as feedback to the CPS or it could be plotted to be observed by a human supervisor if needed.}
  \vspace{-12pt}
\label{Fig:over:solution}
\end{figure}

\responsehighlight{In the following, we represent the set of natural numbers including zero by $\Ne$ and the finite interval of $\Ne$ up to $m$ by $\Ne_m = \{0,1,\ldots,m\}$.}
In this work, we consider monitoring of Cyber-Physical Systems (CPS).
We assume that we have access to some discrete time execution or simulation traces of the CPS. 
\responsehighlight{
We view {\it (execution or simulation) traces} as timed state sequences $\Tc$ = $\Tc_0\Tc_1\Tc_2$ $\ldots$ $\Tc_m$ = $(\tau_0,s_0)$ $(\tau_1,s_1)$ $(\tau_2,s_2)$ $\ldots$ $(\tau_m,s_m)$ 
where for each $k \in \Ne_m$, 
$\tau_k \in \Re_{\geq 0}$ is a time stamp and $s_k \in S$ is a vector containing the values of the state variables of the system at each sampling instance $k$.
For example, for $m=2$, the trace 
$\Tc =  (0,(2,0.34))(0.1,(3,0.356))(0.2,(2,0.36))$
captures the finite time execution of a CPS with two state variables in the vector $s_k$: one ranging over the natural numbers $\Ne$ and the other over the reals $\Re$.
That is, for $k=1$, the state of the system at time $\tau_1 = 0.1$ was $s_1 = (3,0.356) \in \Ne \times \Re$.
} 
We further assume that $\mathcal{S}=(S,d)$ is a generalized quasi-metric space \cite{SedaH08}. 
The existence of metrics is necessary so that distances can be defined for quantitative valuations of the atomic propositions \cite{FainekosP09tcs,Abbas13tecs}.

\responsehighlight{
Throughout the paper, the variable $i$, which ranges over $\Ne$, is used to represent the current simulation step or the current index of the sampling process. 
} 
We assume a fixed sampling period for the monitored system.
Thus, there exists a fixed time period between consecutive time stamps. 
For the fixed time period $\Delta t >0$, for all $i\geq 0$, we have $\tau_{i+1}-\tau_i = \Delta t$ (or equivalently $\tau_i = i \Delta t$). 
As a result, we can simply compute each time stamp $\tau_i$ knowing the trace index (or simulation step) $i$ by multiplication ($\tau_i = i \Delta t$). 
Therefore, we use the trace index (simulation step $i$) as the reference of time. 

The property of interest is stated in Metric Temporal Logic (MTL) with bounded future and unbounded past (\bMITL) for timed state sequences \cite{Thati05entcs}. 
\responsehighlight{
More specifically, at each time $i$, we would like to monitor safety requirements represented as \bMITL 
formulas.} These formulas capture safety properties of the system, such as bounded reactivity, which can be periodically analyzed for violation.  
In our formulation, we use the robust (quantitative) semantics \cite{FainekosP09tcs} that quantify the distance between a given execution trace of a CPS and all the execution traces that violate the property. 
The robustness of a formula $\llbracket\varphi\rrbracket$ with respect to a trace $\Tc$ at time $i$ is a value that measures how far is the trace from the satisfaction/falsification. 
This measure is an extension of boolean values representing satisfaction or falsification which is used in conventional monitoring. 
A positive robustness value means that the trace satisfies the property and a negative robustness means that the specification is not satisfied.

\responsehighlight{
Our goal in this paper is to provide monitoring tools for temporal logic robustness.
We assume that at each time $i$, the CPS outputs its current state $s_i$ along with a finite prediction $s_{i+1}$, $s_{i+2}$, $\ldots$, $s_{i+Hrz}$ of horizon length $Hrz \in \Ne$ (see Fig. \ref{Fig:over:solution}).
The horizon length $Hrz$ will be formally defined in Sec. \ref{sec:rob}; however, informally, it is the required number of samples after time $i$ so that any future requirements in the MTL specification $\phi$ are resolved, i.e., the horizon depends on the structure of the formula $\phi$, $Hrz = hrz(\phi)$.
When dealing with CPS, there exist numerous methods by which such a prediction horizon (forecasting) can be computed \cite{EklundSS05acc,BakirtzisEtAl96tps,MonteiroEtAl09anl}.
}
 
Next, we formally define the main problem presented in this paper.

\begin{prob}[\bMITL Robustness Monitoring]
\responsehighlight{
Given an \bMITL specification $\varphi$, a sampling instance $i$ and an execution trace $\Tc = \Tc_0 \Tc_1 \ldots \Tc_m$ such that $m = i+hrz(\varphi)$, compute the current robustness estimate $\dle \varphi\dri(\Tc,i)$ at time $\tau_i$.
} 
\end{prob}

\responsehighlight{
Intuitively, $\varphi$ represents a system invariant that must hold at every point in the system execution. 
This can also be viewed as testing for the specification robustness $\dle\Box\varphi\dri(\Tc,0)$, where $\Box$ is the operator for ``{\it always in the future}" and $\varphi$ is an arbitrary \bMITL specification. 
However, instead of caring about the satisfaction of the formula at the beginning of the time, we care about the potential of violating $\varphi$ for which we design an on-line monitor.
}

\noindent{\bf Overview of solution and summary of contributions:}
We provide an on-line monitoring approach for computing the robustness of an \bMITL formula with respect to execution traces of a CPS. 
An overview of the solution for the \bMITL on-line monitoring problem appears in Fig. \ref{Fig:over:solution}. 
\responsehighlight{Our method monitors the behavior of a CPS as it executes. 
Our toolbox is also useful for applications where Simulink models are actually used for process monitoring (and not simulation). In addition, it can also be used for code generation for general \bMITL monitors for deployment on actual systems.}
Our method computes the robustness of invariants $\dle \varphi \dri(\Tc,i)$ by storing previous specification robustness values -- if needed -- and by only utilizing a bounded number of pairs of the execution trace \responsehighlight{$\Tc_{Hst}, \ldots, \Tc_{Hrz}$ where $Hst \in \Ne_{i}$ and it will be formally defined in Sec. \ref{sec:rob}}. 
Our monitor uses bounded memory and, in the worst case, it has quadratic time complexity that depends on the magnitude of $Hrz-Hst$.
In principle, our solution for robustness monitoring is inspired by the boolean temporal logic monitoring algorithm in \cite{FinkbeinerK09}.

%% file: prelim.tex
\section{Robustness of Metric Temporal Logic Specifications}
\label{sec:mtl}

In digital control and monitoring of CPS, it is inevitable that physical quantities are measured through a sampling process. 
As mentioned in the Problem Formulation section, when we mention time, we are actually referring to the corresponding sampling index $i$. 
\responsehighlight{
With a slight abuse of notation and under the assumption of constant sampling rate, an execution trace $\Tc$ can also be represented by a function $s: \Ne_{i+Hrz} \rightarrow S$.
The view of the sequence $s_0 s_1 \ldots s_{i+Hrz}$ as a function $s$ simplifies the presentation of the robust semantics for MTL.
}

Using a metric $d$ \cite{SedaH08}, we can define a distance function that captures \responsehighlight{how far away a point $x \in X$ is from a set $S \subseteq X$}. 
Intuitively, the distance function assigns positive values when $x$ is in the set $S$ and negative values when $x$ is outside the set $S$. \responsehighlight{The metric
$d$ must be at least a generalized quasi-metric as described in \cite{Abbas13tecs} which also includes the case where $d$ is a metric as it was introduced in \cite{FainekosP09tcs}.}

\begin{definition}[Signed Distance] Let $x \in X$ be a point, $S \subseteq X$ be a set and $d$ be a metric. Then, we define the Signed Distance from $x$ to $S$ to be 
\[ \mathbf{Dist}_d(x,S) := \left\{ \begin{array}{ll}
-\inf\{d(x,y)\;|\;y \in S\} & \mbox{ if } x \not \in S \\
\inf\{d(x,y)\;|\;y \in X \backslash S \} & \mbox{ if } x \in S \\
\end{array} \right. \]
\responsehighlight{where inf is the infimum.}
\end{definition}

Metric Temporal Logic (MTL) was introduced by Koymans \cite{Koymans90} to reason about the quantitative timing properties of boolean signals. 
In this paper, we use the standard fragment of MTL with bounded future, but also we allow the use of past time operators.

\begin{definition}[\bMITL Syntax] 
Let $AP$ be the set of atomic propositions and $\mathcal{I}$ be any non-empty interval of $\mathbb{N}$, and $ \overline{\mathcal{I}}$ be any non-empty interval of $\mathbb{N}\cup\{+\infty\}$. 
The set \bMITL formulas is inductively defined as $\varphi \; ::=\; \top \; | \; p \; | \; \neg \varphi \; | \; \psi \vee \varphi \; | \; \psi \mathcal{U}_{\mathcal{I}} \varphi \; | \; \psi \mathcal{S}_{\overline{\mathcal{I}}} \varphi $ where $p \in AP$ and $\top$ stands for \emph{true}.
\end{definition}

Note that we use the number of samples to represent the time interval constraints of temporal operators. 
For example assume that $\Delta t = 0.1$, then the MTL formula $\Diamond_{[0,0.5]}a$ where the timing constraints are over time is instead represented by $\Diamond_{[0,5]}a$ in \bMITL. 

The propositional operators conjunction ($\wedge$) and implication ($\rightarrow$) are defined the usual way.
All other bounded future temporal operators can be syntactically defined using Until ($\Uc_\Ic$), where $\medcirc$ (Next), $\Diamond$ (Eventually), and $\Box$ (Always) are defined as $\medcirc\varphi\equiv\top\Uc_{[1,1]}\varphi$, $\Diamond_\Ic\varphi\equiv\top\Uc_\Ic\varphi$, and $\Box_\Ic\varphi\equiv\neg\Diamond_\Ic\neg\varphi$ respectively. \responsehighlight{The intuitive meaning of the $\psi \mathcal{U}_{[a,b]} \varphi$ operator at sampling time $i$ is a follows: $\psi$ has to hold at least until $\varphi$ becomes true within the time interval of $[i+a,i+b]$ in the future}.
Similarly, all other bounded/unbounded past temporal operators can be defined using Since ($\Sc_{\overline{\Ic}}$), where $\odot$ (Previous), $\Diamonddot$ (Eventually in the past), and $\boxdot$ (Always in the past) are defined as $\odot\varphi\equiv\top\Sc_{[1,1]}\varphi$, $\Diamonddot_{\overline{\Ic}}\varphi\equiv\top\Sc_{\overline{\Ic}}\varphi$, and $\boxdot_{\overline{\Ic}}\varphi\equiv\neg\Diamonddot_{\overline{\Ic}}\neg\varphi$ respectively. \responsehighlight{The intuitive meaning of the $\psi \mathcal{S}_{[a,b]} \varphi$ operator at sampling time $i$ is as follows: since $\varphi$ becomes true within the interval $[i-b,i-a]$ in the past, $\psi$ must hold till now (current time $i$).}

\bMITL can state requirements over the observable trajectories of a CPS. 
In order to capture these requirements, each predicate $p\in AP$ is mapped to a subset of the metric space $X$. 
We use an observation map $\Oc$ to interpret each predicate $p\in AP$. 
In other words, the observation map is defined as $\Oc : AP \rightarrow P(X)$ such that for each $p \in AP$ the corresponding set is $\Oc(p)$. 
Here, $P(S)$ denotes the powerset of a set $S$. 
We define the robust valuation of an \bMITL formula $\varphi$ over a \responsehighlight{trace} $s$ as follows \cite{FainekosP06fates}.

\begin{definition}[\bMITL Robustness Semantics] 
Let $s$ be a \responsehighlight{trace} $s:\mathbb{N} \rightarrow X$, and $\Oc$ be an observation map $\mathcal{O} : AP \rightarrow P(X)$, then the robust semantics of any formula $\varphi \in$ \bMITL with respect to $s$ is recursively defined as:
\begin{align*}
  \dle \top \dri (s,i) & := +\infty \\
\dle p \dri(s,i) & :=\mathbf{Dist}_d(\responsehighlight{s(i)},\mathcal{O}(p)) \\
\llbracket\neg\varphi\rrbracket(s,i) & :=-\llbracket\varphi\rrbracket(s,i) \\
\llbracket\psi\vee\varphi\rrbracket(s,i) & :=\llbracket\psi\rrbracket(s,i)\sqcup\llbracket\varphi\rrbracket(s,i) \\
\llbracket\psi\mathcal{U}_{[l,u]}\varphi\rrbracket(s,i) & :=  \sideset{}{_{j=i+l}^{i+u}}\bigsqcup\bigg(\llbracket\varphi\rrbracket(s,j)\sqcap\sideset{}{_{k=i}^{j-1}}\bigsqcap\llbracket\psi\rrbracket(s,k)\bigg) \\
\llbracket\psi\mathcal{S}_{[l',u'\rangle}\varphi\rrbracket(s,i) & := \sideset{}{_{j=max\{0,i-u'\}}^{i-l'}}\bigsqcup\bigg(\llbracket\varphi\rrbracket(s,j)\sqcap\sideset{}{_{k=j+1}^{i}}\bigsqcap\llbracket\psi\rrbracket(s,k)\bigg)
\end{align*}
where $\sqcup$ stands for max, $\sqcap$ stands for min, $p\in AP$, $l,u,l' \in \Ne$ and $u' \in \Ne \cup \{\infty\}$.
Furthermore, the symbol $\rangle$ in $\mathcal{S}_{[l',u'\rangle}$ will be ) when $u'=+\infty$ and $]$ when $u'\ne+\infty$. 
\end{definition}

We should point out that we use the extended definition of maximum ($\sqcup$) and minimum ($\sqcap$), \responsehighlight{ with slight abuse of notation, we let $\max(\emptyset) = -\infty $ and $\min(\emptyset) = +\infty$.  i.e., over empty sets we treat min and max as infimum and supremum, respectively. 
For exact definition of infimum and supremum see \cite{Davey02lattice}}.

%% file: finitehorizon.tex
\section{Robustness Monitoring of \bMITL}
\label{sec:rob}
\subsection{Finite horizon and history of \bMITL}
For each \bMITL formula $\psi$ we define the finite horizon $hrz(\psi)$ as the number of samples we need to consider in the future. \responsehighlight{In MTL, the satisfaction of the formula depends on what will happen in the future. In bounded MTL, the finite horizon $hrz(\psi)$ is the number of steps (samples) which we need to consider in the future in order to evaluate the formula $\psi$ at the current time $i$.} In other words, $hrz(\psi)$ is the number of steps into the future for which the truth value of the sub-formula $\psi$ depends on \cite{FinkbeinerK09}. Similarly, we define the finite history $hst(\psi)$ of $\psi$ as the number of samples we need to look into the past. That is, the number of steps in the past for which the truth value of the sub-formula $\psi$ depends on. \responsehighlight{Intuitively, the $hst(\psi)$ is the size of the history we need to consider in order to keep track of what happened in the past to evaluate the formula $\psi$ at present time.} The finite horizon and the history can be defined recursively. We define $hrz(\psi)$ (similar to $h(\psi)$ in \cite{FinkbeinerK09}) and we add the recursive definition of $hst(\psi)$ in the following:
\vspace{-4pt}
\begin{flushleft}
\begin{tabular} {llll}
$hrz(p)=0$ & $hst(p)=0$ \\ $hrz(\neg\psi)=hrz(\psi)$ & $hst(\neg\psi)=hst(\psi)$ \\[2pt] $hrz(\psi$ {\bf OP }$\varphi)=max\{hrz(\psi),hrz(\varphi)\}$ & $hst(\psi$ {\bf OP }$\varphi)=max\{hst(\psi),hst(\varphi)\}$ \\[2pt]
$hrz(\psi\Uc_{[l,u]}\varphi)=max\{hrz(\psi)+u-1 , hrz(\varphi)+u\}$ \hspace{15pt} & $hst(\psi\Uc_{[l,u]}\varphi)=max\{hst(\psi) , hst(\varphi)\}$ \\[2pt]
$hrz(\psi\Sc_{[l',u'\rangle}\varphi)=max\{hrz(\psi),hrz(\varphi)\}$    \\[2pt] 
\end{tabular}
$hst(\psi\Sc_{[l',u'\rangle}\varphi)=\left\{ \begin{array}{ll}
max\{hst(\psi)+u'-1 , hst(\varphi)+u'\} & \mbox{ if }  u'\ne+\infty \\
max\{hst(\psi)+l'-1 , hst(\varphi)+l'\} & \mbox{ if }  u'=+\infty \\
\end{array} \right.$
\end{flushleft}
where $p\in AP$.
Here, ${\bf OP }$ is any binary operator in propositional logic, and $\psi,\varphi$ are \bMITL formulas. For the unbounded $\Sc_{[0,+\infty)}$ operator, the computation of finite history is more involved and needs more explanation. \responsehighlight{Namely, we need to restate the dynamic programming algorithm for monitoring a sub-formula $\psi\Sc_{[0,+\infty)}\varphi$ based on the following works \cite{Rosu01,HavelundR04}}.
According to the robustness semantics, the robustness of $\psi\Sc_{[0,+\infty)}\varphi$ at time $i$ is as follows:
 
\[ \llbracket\psi\mathcal{S}_{[0,+\infty)}\varphi\rrbracket(s,i)= \sideset{}{_{j=0}^{i}}\bigsqcup\bigg(\llbracket\varphi\rrbracket(s,j)\sqcap\sideset{}{_{k=j+1}^{i}}\bigsqcap\llbracket\psi\rrbracket(s,k)\bigg)\]
also robustness of $\psi\Sc_{[0,+\infty)}\varphi$ at time $i-1$ is
\[\llbracket\psi\mathcal{S}_{[0,+\infty)}\varphi\rrbracket(s,i-1)= \sideset{}{_{j=0}^{i-1}}\bigsqcup\bigg(\llbracket\varphi\rrbracket(s,j)\sqcap\sideset{}{_{k=j+1}^{i-1}}\bigsqcap\llbracket\psi\rrbracket(s,k)\bigg)\]

Thus, we can rewrite $\llbracket\psi\mathcal{S}_{[0,+\infty)}\varphi\rrbracket(s,i)$ as

\begin{gather*}
\llbracket\psi\mathcal{S}_{[0,+\infty)}\varphi\rrbracket(s,i)=\llbracket\varphi\rrbracket(s,i)\sqcup\Bigg(\llbracket\psi\rrbracket(s,i)\sqcap\bigg(\sideset{}{_{j=0}^{i-1}}\bigsqcup\Big(\llbracket\varphi\rrbracket(s,j)\sqcap\sideset{}{_{k=j+1}^{i-1}}\bigsqcap\llbracket\psi\rrbracket(s,k)\Big)\bigg)\Bigg)= \\
 = \llbracket\varphi\rrbracket(s,i)\sqcup\Bigg(\llbracket\psi\rrbracket(s,i)\sqcap\bigg(\llbracket\psi\mathcal{S}_{[0,+\infty)}\varphi\rrbracket(s,i-1)\bigg)\Bigg)
\end{gather*}

Therefore, similar to \cite{HavelundR04} we recursively update the robustness of $\psi\Sc_{[0,+\infty)}\varphi$ at the current time $i$ and save it in a variable called ``$Pre$"  to reuse it for the computation of the next time step (see \cite{HavelundR04} for more details).
As a result, when we have an unbounded past time operator, we do not need the full history table. However, if the formula contains a nested future time operator, we need to extend the history to be long enough to contain the actual values. In other words, although for unbounded past time operators we do not need the whole history table, we should still extend the history to be able to store the actual simulation values (not the predicted values) in ``$Pre$".
 

\subsection{Robustness Computation Algorithm}

\begin{table}[t]
  \centering
  \caption{Pre Vector and Robustness Table  }
  \fontsize{8.2}{10}
  \begin{tabular}{ c||c|c||c|c|c|c|c| }
  \hline
  Pre[$k$] &$T_{k,j}$& column $j\Rightarrow$ & -2 & -1 & 0 & 1 & 2\\ \hline
   & row $k\Downarrow$&Time$(i)$ & $i-2$ & $i-1$ & \hspace{5pt} $i$ \hspace{5pt} & $i+1$ & $i+2$ \\ \hline\hline
  \cellcolor{gray!50}&$\psi_1=\varphi$& $\psi_2\wedge\psi_3$ & $\llbracket\varphi\rrbracket(s,i-2)$ & $\llbracket\varphi\rrbracket(s,i-1)$ & $\llbracket\varphi\rrbracket(s,i)$ & $\llbracket\varphi\rrbracket(s,i+1)$ & $\llbracket\varphi\rrbracket(s,i+2)$ \\ \hline
  \cellcolor{gray!50}&$\psi_2$& $\Box_{[1,2]}q$ & $\llbracket\psi_2\rrbracket(s,i-2)$ &$\llbracket\psi_2\rrbracket(s,i-1)$ &$\llbracket\psi_2\rrbracket(s,i)$ &$\llbracket\psi_2\rrbracket(s,i+1)$ & $\llbracket\psi_2\rrbracket(s,i+2)$ \\ \hline
  $\llbracket\psi_3\rrbracket(s,i-3)$&$\psi_3$& $\boxdot_{[0,+\infty)}p$ & $\llbracket\psi_3\rrbracket(s,i-2)$ & $\llbracket\psi_3\rrbracket(s,i-1)$ &$\llbracket\psi_3\rrbracket(s,i)$ & $\llbracket\psi_3\rrbracket(s,i+1)$ & $\llbracket\psi_3\rrbracket(s,i+2)$ \\ \hline
  \cellcolor{gray!50}&$\psi_4$& $p$ & $\llbracket\psi_4\rrbracket(s,i-2)$ & $\llbracket\psi_4\rrbracket(s,i-1)$ &$\llbracket\psi_4\rrbracket(s,i)$  & $\llbracket\psi_4\rrbracket(s,i+1)$ & $\llbracket\psi_4\rrbracket(s,i+2)$ \\ \hline
  \cellcolor{gray!50}&$\psi_5$& $q$ & $\llbracket\psi_5\rrbracket(s,i-2)$ & $\llbracket\psi_5\rrbracket(s,i-1)$ &$\llbracket\psi_5\rrbracket(s,i)$ & $\llbracket\psi_5\rrbracket(s,i+1)$ & $\llbracket\psi_5\rrbracket(s,i+2)$ \\
  \hline
  \end{tabular}
  \label{tab:PreRobTable}%
    \vspace{-12pt}
\end{table}%

\begin{table}[b]
  \centering
    \vspace{-12pt}
  \caption{Robustness Computation of each table entries (Gray cells are unused) }
  \fontsize{9}{10}
  \begin{tabular}{ c||c|c|c|c|c| }
  \hline
  $T_{k,j}$ & $i-2$ & $i-1$ & $i$ & $i+1$ & $i+2$ \\ \hline
  $k\Downarrow$,$j\Rightarrow$ & $j=-2$ & $j=-1$ & $j=0$ & $j=1$ & $j=2$\\ \hline\hline
  Pre[1]\cellcolor{gray!50} &$T_{2,-2}\sqcap T_{3,-2}$&$T_{2,-1}\sqcap T_{3,-1}$&$T_{2,0}\sqcap T_{3,0}$&$T_{2,1}\sqcap T_{3,1}$& $T_{2,2}\sqcap T_{3,2}$\\ \hline
Pre[2]\cellcolor{gray!50} & $T_{5,-1}\sqcap T_{5,0}$ & $T_{5,0}\sqcap T_{5,1}$ &$T_{5,1}\sqcap T_{5,2}$ & $T_{5,2}$ & $+\infty$ \\ \hline
  Pre[3] & Pre[3]$\sqcap T_{4,-2}$ & $T_{3,-2}\sqcap T_{4,-1}$ & $T_{3,-1}\sqcap T_{4,0}$ & $T_{3,0}\sqcap T_{4,1}$ & $T_{3,1}\sqcap T_{4,2}$ \\ \hline
  Pre[4]\cellcolor{gray!50} &{\bf Dist}$_d(s_{i-2} ,\mathcal{O}(p))$&{\bf Dist}$_d(s_{i-1} ,\mathcal{O}(p))$ &{\bf Dist}$_d(s_{i} ,\mathcal{O}(p))$& {\bf Dist}$_d(s_{i+1} ,\mathcal{O}(p))$&{\bf Dist}$_d(s_{i+2} ,\mathcal{O}(p))$ \\ \hline
  Pre[5]\cellcolor{gray!50} &{\bf Dist}$_d(s_{i-2} ,\mathcal{O}(q))$&{\bf Dist}$_d(s_{i-1} ,\mathcal{O}(q))$ &{\bf Dist}$_d(s_{i} ,\mathcal{O}(q))$& {\bf Dist}$_d(s_{i+1} ,\mathcal{O}(q))$&{\bf Dist}$_d(s_{i+2} ,\mathcal{O}(q))$\\
  \hline
  \end{tabular}
  \label{tab:CompRob}%
  
\end{table}%

\begin{algorithm}[t]
\fontsize{7.5}{7.5}
\caption{On-Line Monitor}
\responsehighlight{{\bf Input}: $\varphi$, $s'_i=s_is_{i+1}\ldots s_{i+Hrz}$, $\genMet$, $\Oc$; 
{\bf Global variables:} $T$, $Pre$; 
{\bf Output}: $T_{1,0}(robustness$ $value)$.}
\label{alg:on-line}
\begin{multicols*}{2}
 \fontsize{8.5}{8.5}
 \hspace {8pt}{\bf   procedure  }{\sc Monitor}($\varphi, \responsehighlight{s'_i}, \genMet, \Oc$)
\begin{algorithmic}[1]
\For{$k\gets 1\mbox{ to }|\varphi|$}
\State  $Pre(k)\gets T_{k,(-Hst+hst(\varphi_k))}$
\EndFor
 \For{$j\gets 1-Hst\mbox{ to }Hrz$}
 \For{$k\gets 1\mbox{ to }|\varphi|$}
 \If{$\varphi_k=p\in AP$}
 	\State $T_{k,j-1}\gets T_{k,j}$
 \EndIf
 \EndFor 
 \EndFor
 \For{$k\leftarrow |\varphi|\mbox{ down to }1$}
\If{\responsehighlight{$\varphi_k=\varphi_m\mathcal{S}_{[l',u'\rangle}\varphi_n$}}
   \For{\yhl{$j\leftarrow -Hst+hst(\varphi_k)\mbox{ to }Hrz$}}
\State\hspace{-12pt}    \responsehighlight{ $T_{k,j}\leftarrow CR(\varphi_k,j, s'_i, \genMet,\Oc)$}
\EndFor
\Else  
  \For{\yhl{$j\leftarrow Hrz\mbox{ down to }-Hst+hst(\varphi_k)$}}
\State\hspace{-12pt}   \responsehighlight{$T_{k,j}\leftarrow CR(\varphi_k,j, s'_i, \genMet,\Oc)$}
\EndFor
\EndIf
 \EndFor
\State\Return$T_{1,0}$
\end{algorithmic}
\hspace {5pt}{\bf   end procedure}
\end{multicols*}
\end{algorithm}
\begin{algorithm}[t]
\fontsize{7.5}{7.5}
\caption{Robustness Computation (CR)}
\responsehighlight{{\bf Input}: $\varphi_k$, $j$, $s'_i=s_is_{i+1}\ldots s_{i+Hrz}$, $\genMet$, $\Oc$;
{\bf Global variables:} $T$, $Pre$;
{\bf Output}: $T_{k,j}$.}
\label{alg:comprob}
\begin{multicols}{2}
\hspace {8pt}{\bf   procedure  }{\sc CR}\responsehighlight{($\varphi_k, j, s'_i, \genMet, \Oc$)}
\begin{algorithmic}[1]
\If {$\varphi_k=\top$} $T_{k,j}\leftarrow+\infty$
\ElsIf {$\varphi_k=p\in AP$}
\If{$j>=0$}
\State	  \responsehighlight{$T_{k,j}\leftarrow Dist_d(s_{i+j},\mathcal{O}(p))$}
\EndIf
\ElsIf{$\varphi_k=\neg\varphi_m$}
\State $T_{k,j}\leftarrow -T_{m,j}$
\ElsIf{$\varphi_k=\varphi_m\vee\varphi_n$}
\State $T_{k,j}\leftarrow T_{m,j}\sqcup T_{n,j}$
\ElsIf{$\varphi_m\mathcal{U}_{[l,u]}\varphi_n$}
\If{$j+l\le Hrz$}
\State $tmp_{min}\leftarrow\bigsqcap_{j\le j'< j+l} T_{m,j'}$
\State $T_{k,j}=-\infty$
\For   {$j'\leftarrow j+l\mbox{ to }min\{Hrz,j+u\}$}
\State 	 	  $T_{k,j}\leftarrow T_{k,j}\sqcup(tmp_{min}\sqcap T_{n,j'})$
\State 	 	  $tmp_{min}=tmp_{min}\sqcap T_{m,j'}$
\EndFor
\Else {}
\State$T_{k,j}=-\infty$
\EndIf
\ElsIf{$\varphi_m\mathcal{S}_{[l',u'\rangle}\varphi_n$}
\State  $tmp_{min}\leftarrow\bigsqcap_{j-l'< j'\le j} T_{m,j'}$
\If{$u'\ne +\infty$}
\State $T_{k,j}=-\infty$
\For{\yhl{$j'\leftarrow j-l'\mbox{ down to }j-u'$}}
\State  $T_{k,j}\leftarrow T_{k,j}\sqcup(tmp_{min}\sqcap T_{n,j'})$
\State	  $tmp_{min} = tmp_{min}\sqcap T_{m,j'}$
\EndFor
\Else {}
\If{\yhl{$j=-Hst+hst(\varphi_k)$}}
\State $tmp_{\Sc} \leftarrow Pre[k]\sqcap T_{m,j}$
\Else {}
\State $tmp_{\Sc} \leftarrow T_{k,j-1}\sqcap T_{m,j}$
\EndIf
\State $T_{k,j} \leftarrow (T_{n,j-l'}\sqcap tmp_{min}) \sqcup tmp_{\Sc} $
\EndIf
\EndIf 
\State\Return{$T_{k,j}$}
\end{algorithmic}
\hspace {5pt}{\bf   end procedure}
\end{multicols}
\vspace*{-10pt}
\end{algorithm}

For each \bMITL formula $\varphi$ we construct a table called {\bf Robustness Table} with width of \responsehighlight{$Hst+1+Hrz$, where $Hrz=hrz(\varphi)$ is the finite horizon of the specification formula $\varphi$, and, $Hst=Hrz+hst(\varphi)$, where $hst(\varphi)$ is the finite history of the specification $\varphi$. $Hst$ is extended conservatively due to the fact that ``$Pre$" value can only store the robustness values corresponding to the actual simulation. The height of the robustness table is the size of the formula $\varphi$ ($|\varphi|$), where $|\varphi|$ is the number of sub-formulas of $\varphi$ including itself.}
For example, assume we have a formula $\varphi=\boxdot_{[0,+\infty)}p\wedge\Box_{[1,2]}q$ and \responsehighlight{we intend to compute $\dle \varphi\dri(\Tc,i)$ at each time $i$}. \responsehighlight{In formula $\varphi$, $Hst=2$ and $Hrz=2$}.
Since $\varphi$ has unbounded past-time operators, it needs the $Pre$ vector as well as the Robustness Table. 
The $Pre$ vector appended to the Robustness Table is presented in Table \ref{tab:PreRobTable}. \responsehighlight{ In particular, the $Pre$ vector contains the value of past sub-formulas from the beginning of the time up to the current time.}

\responsehighlight{ In the following, we explain how the values of Table \ref{tab:CompRob}, the robustness table,  are computed using Algorithms \ref{alg:on-line} and \ref{alg:comprob}. In order to make our algorithms more readable, we used a vector to show the CPS output $s_{i}$, $s_{i+1}$, $\ldots$, $s_{i+Hrz}$ to the monitoring (see Fig. \ref{Fig:over:solution}). We define a vector $s'_i=s_is_{i+1}\ldots s_{i+Hrz}$ which appends current state $s_i$ with predictions $s_{i+1}$, $s_{i+2}$, $\ldots$, $s_{i+Hrz}$.} In Table \ref{tab:PreRobTable}, $i$ is the current simulation step which corresponds to column 0. 
At each simulation step $i$, for each unbounded past time sub-formula $\phi$, we first save the values of the column $-Hst+hst(\phi)$ in the $Pre$ vector \responsehighlight{(Algorithm \ref{alg:on-line} lines 1-3) since the column $-Hst+hst(\phi)$ contains the robustness value of $\phi$ from the beginning of the simulation. We need the $Pre$ vector to compute the robustness of $\phi$ at the next sampling time using the dynamic programming method. In the above example, 
for $\boxdot_{[0,+\infty)}p$ the value at column $-2$ is saved in $Pre$ to be used during robustness computation. 
Then, we shift all the \yhl{robustness table entries of the predicates by one position to the left} (Algorithm \ref{alg:on-line}, \yhl{lines 4-10}). Then the loop (Algorithm \ref{alg:on-line}, \yhl{lines 11-21}) recursively calls Algorithm \ref{alg:comprob} to fill the robustness table for each sub-formula from bottom to top.}

\responsehighlight{Each call of Algorithm \ref{alg:comprob} (CR) computes each table entry $T_{k,j}$ (see tables \ref{tab:PreRobTable},\ref{tab:CompRob}) where column $j$ is the horizon/history index and row $k$ is the sub-formula index. For past sub-formulas the table entries are computed from left to right (Algorithm \ref{alg:on-line}, \yhl{lines 13-15}), and for future sub-formulas the table entries are computed from right to left (Algorithm \ref{alg:on-line}, \yhl{lines 17-19}).
New values for predicates (according to execution traces) will be placed in column $0$ and the predicted values of the predicates will be saved in columns 1 to $Hrz$ (Algorithm \ref{alg:comprob}, lines 2-5). Table \ref{tab:CompRob} shows the updates of predicate values in rows 4, and 5 which correspond to Algorithm \ref{alg:comprob}, line 4.}
 
\responsehighlight{In the following, we explain how the CR Algorithm \ref{alg:comprob} computes the MTL robustness values for three different cases of MTL:\\
{\bf Case 1 (Lines 10-20):} The robustness of bounded future temporal sub-formulas with interval $[l,u]$ at each column $j$ is computed given the values of its operands for columns $j$ up-to $min\{j+u,Hrz\}$ (Line 14). For example, this case is used in Table \ref{tab:CompRob} to compute the robustness of sub-formula $\psi_2=\Box_{[1,2]}q$ from right to left. Case 1 in CR Algorithm is similar to the DP-TALIRO algorithm \cite{FainekosSUY12acc}.\\
{\bf Case 2 (Lines 23-28):} The robustness of bounded past temporal sub-formulas with interval $[l',u']$ at each column $j$ is computed given the values of its operands for columns\yhl{ $j$ down-to $j-u'$ (Line 25).}\\ 
{\bf Case 3 (Lines 30-36):} The robustness of unbounded past temporal sub-formulas with interval $[l',+\infty)$ for column $j$ is computed using the stored value in column $j-1$ in dynamic programming fashion \yhl{(Line 33)} and using the $Pre$ vector \yhl{(Line 31)}. For example, Case 3 is used to compute the robustness of $\psi_3=\boxdot_{[0,+\infty]}p$ using $Pre[3]$ from left to right in Table \ref{tab:CompRob}. 
}

Finally, we update table entries for the top row which corresponds to $\psi_1=\varphi$.
Since its corresponding operator $\wedge$ is propositional (Algorithm \ref{alg:comprob} Lines 6-9), we can update its value from any direction. The high level explanation of Algorithm \ref{alg:on-line} is described as follows:

\begin{enumerate}
\vspace{-6pt}
\item Store values of column $-Hst+hst(\phi_k)$ for each unbounded past sub-formula $\phi_k$ in $Pre[k]$ and shift the \yhl{table entries of predicates one to the left (Lines 1-10)}.
\item For each row $i$ from $|\varphi|$ to 1 compute the robustness values according to:
\begin{enumerate}
\item If $\varphi_i$ is a future temporal operator, for each column $j$ from $Hrz$ \yhl{down to $-Hst+hst(\varphi_i)$}, update table entry $T_{i,j}$ using Algorithm \ref{alg:comprob}.
\item If $\varphi_i$ is a past temporal operator, for each column $j$ from \yhl{$-Hst+hst(\varphi_i)$} up to $Hrz$ update table entry $T_{i,j}$ using Algorithm \ref{alg:comprob}.
\end{enumerate}
\item\responsehighlight{ Return the robustness ($T_{1,0}$).}
\end{enumerate}
\yhl{We provided the proof of this section in Appendix.}

%% file: experiments.tex
\section{Experimental Analysis and Case Studies}

\subsection{Runtime Overhead}

First, we measure the overhead of the proposed monitoring framework on a slightly modified version of the Automatic Transmission (AT) model provided by Mathworks as a Simulink demo\footnote{Available at: \url{http://www.mathworks.com/help/simulink/examples/modeling-an-automatic-transmission-controller.html}}.
The experiments were conducted on a Windows 7, Intel Core2 Quad (2.99 GHz) with 8 GB RAM. 

The physical model of the AT system has two continuous (real-valued) state variables which are also its monitored outputs: the speed of the engine $\omega$ and the speed of the vehicle $v$.
The model includes an automatic transmission controller that exhibits both continuous and discrete behavior.
It is a typical CPS model and specifications over both boolean and continuous variables can be formalized.
However, since the valuation of the robustness of predicates over continuous state variables is computationally more expensive than a boolean valuation, we consider only specifications over continuous state variables for the impact analysis.

We introduce our \bMITL monitoring block in the AT model and test the performance over a set of specifications. 
\responsehighlight{In order to test the runtime overhead of our work, we artificially generate 30 different \bMITL formulas based on typical critical safety formulas to show that the runtime overhead depends on both of the size of the formula and the horizon/history. We test our method for 100 runs of monitoring algorithm for each specification (formula), and for each run we use 100 simulation steps. Then, we compute the mean and variance of the overhead for each simulation step which is the execution time of Algorithm \ref{alg:on-line} in Table \ref{tab:overheadRes}. In this table,}
the overhead is measured on specifications that contain either nested Until operators (U columns) or nested Eventually operators (E columns).\\
\responsehighlight{We generate 30 formulas according to} the following templates:

\begin{itemize}
\vspace{-4pt}
\item{\bf E formulas:} $\phi_n(H)=p_j\longrightarrow\psi_n(H/n)$

where $H\in\Ne$ is the finite horizon of the formula. 
In Table \ref{tab:overheadRes}, we used 1,000, 2,000 and 10,000 for the size of the horizon. Here, $p_j$ is an arbitrary predicate and $\psi_n(H/n)$ is defined recursively as follows:

\hspace{10pt} $\psi_1(h)=\Diamond_{[0,h]}p_k$
and
$\psi_n(h)=\Diamond_{[0,h]}(p_l\wedge\psi_{n-1}(h))$, for $1<n\le 9 $ 

where $h=H/n$, i.e., the finite horizon $H$ divided by the number of nested sub-formulas $n$ and $p_k, p_l$ are arbitrary predicates.

\item {\bf U formulas:} $\phi_n(H)=p_j\longrightarrow\psi_n(H/n)$

where $H \in \Ne$ is the finite horizon of the formula. 
In Table \ref{tab:overheadRes}, we used 1,000, 2,000 and 10,000 for the size of the horizon of $H$. Here, $p_j$ is an arbitrary predicate and $\psi_n(H/n)$ is defined recursively as follows:

\hspace{10pt} $\psi_1(h)=p_k\Un_{[0,h]}p_l$
and
$\psi_n(h)=p_m\Un_{[0,h]}(p_n\wedge\psi_{n-1}(Y))$, for $1<n\le 9 $

where $h=H/n$ and $p_k, p_l, p_m, p_m$ are arbitrary predicates.
\end{itemize}

\responsehighlight{As illustrated in Table \ref{tab:overheadRes}, the computational complexity of the monitoring algorithm is closely related to the horizon and history size.}
\responsehighlight{Since the algorithm's complexity is of order $O(n^2)$ where $n$ is  the horizon/history, the added overhead (in worst case execution) is quadratic in terms of the size of the horizon for some formulas in Table \ref{tab:overheadRes} (like $\phi_1(H)$).}
Moreover, in most cases, the impact of the number of nested temporal operators is not significant compared to the size of horizon/history windows.
From Table \ref{tab:overheadRes}, we notice that when the horizon and history size is less than \responsehighlight{2,000}, the overhead for each simulation step is negligible with our prototype implementation.
Furthermore, for most practical reactivity requirements, it is quite unlikely that even a window size of \responsehighlight{2,000} sampling points is necessary. 
Therefore, the method could be utilized in real world monitoring applications.

\begin{table}[t]
  \centering
  \caption{The overhead on each simulation step on the Automatic Transmission model with specifications of increasing length. Table entries are in milliseconds.  }
    \begin{tabular}{p{1.0cm}p{0.7cm}p{0.9cm}p{0.7cm}p{0.9cm}p{0.7cm}p{0.9cm}p{0.7cm}p{0.9cm}p{0.7cm}p{0.9cm}p{0.7cm}p{0.9cm}}
    \toprule         
    \#    & \multicolumn{4}{c}{H=1,000} & \multicolumn{4}{c}{H=2,000} & \multicolumn{4}{c}{H=10,000} \\
    \midrule
           & \multicolumn{2}{l} E     &\multicolumn{2}{l} U     &\multicolumn{2}{l} E     & \multicolumn{2}{l}U     & \multicolumn{2}{l}E     & \multicolumn{2}{l} U \\ \hline
    \multicolumn{1}{l}{}      & Mean  & Var. & Mean  & Var. & Mean  & Var. & Mean  & Var. & Mean  & Var. & Mean  & Var.  \\ \hline           
    \multicolumn{1}{l}{$\phi_1(H)$}      & 2.39  & 0.00 & 4.83  & 0.00  & 8.03  & 0.00 & 15.8  & 0.001  & 188.8  & 0.001  & 358.5 & 0.036 \\
    \multicolumn{1}{l}{$\phi_3(H)$}     & 4.24  & 0.00  & 7.5  & 0.001  & 12.7 & 0.00 & 25.09  & 0.005  & 314.4 & 0.01   & 599  & 0.665\\
    \multicolumn{1}{l}{$\phi_5(H)$}     & 4.66  & 0.00  & 8.36  & 0.001   & 14.01 & 0.00 & 27.8  & 0.005  & 309.2  & 0.077  & 650 & 0.014 \\
    \multicolumn{1}{l}{$\phi_7(H)$}      & 4.95  & 0.00  & 8.94  & 0.00  & 14.83 & 0.00 & 29.33  & 0.006  & 311  & 0.013  & 674.2 & 0.033 \\
    \multicolumn{1}{l}{$\phi_9(H)$}      & 5.23  & 0.00  & 9.46  & 0.001  & 15.4 & 0.001 & 30.56  & 0.007  & 317.5  & 0.011  & 683.5 & 0.698 \\
    \bottomrule
    \end{tabular}%
  \label{tab:overheadRes}%
  \vspace{-10pt}
\end{table}%
\vspace{-10pt}

%% file: casestudy.tex
\subsection{Case Study}

In the following, we utilize the monitoring method on an industrial size high-fidelity engine model. The model is part of the SimuQuest Enginuity \cite{Simuquest:Online} Matlab/Simulink tool package. The Enginuity tool package includes a library of modules for engine component blocks. It also includes pre-assembled models for standard engine configurations. In this work, we use the Port Fuel Injected (PFI) spark ignition, 4 cylinder inline engine configuration. It models the effects of combustion from first physics principles on a cylinder-by-cylinder basis, while also including regression models for particularly complex physical phenomena. The model includes a tire-model, brake system model, and a drive train model (including final drive, torque converter and transmission). The input to the system is the throttle schedule. The output is the normalized air-to-fuel(A/F) ratio. Simulink reports that this is a 56 state model. Note that this number represents only the visible
states. It is possible that more states are present in the blackbox s-functions which are not accessible. This is a high dimensional non-linear system for which reachability analysis is very difficult. It also includes lookup tables, non-linear
components, and inputs that affect the switching guards. 

\begin{figure}[tbh]
\begin{center}
\includegraphics[width=9cm]{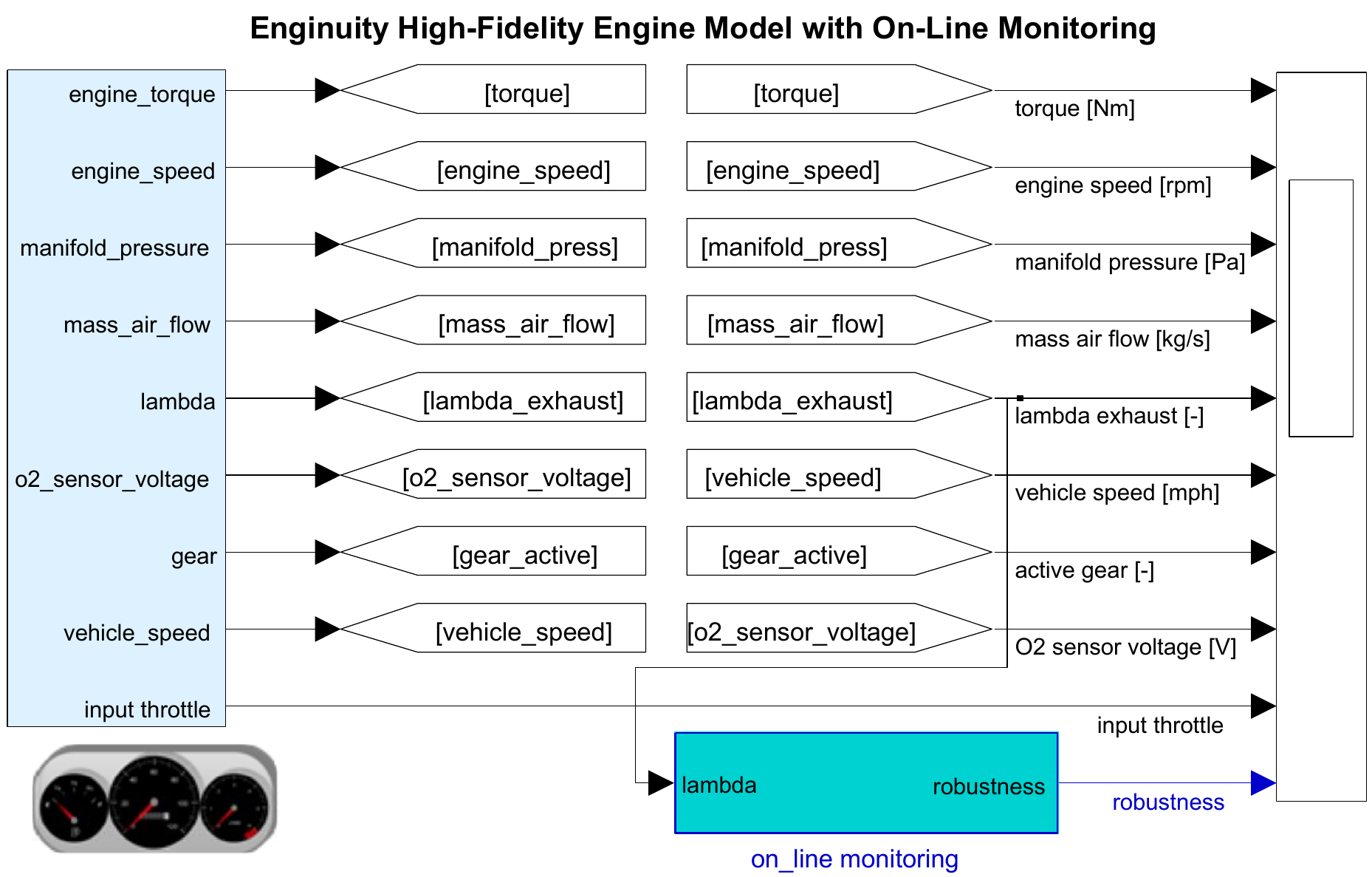} 
\end{center}
\vspace{-0.5cm}
\caption{SimuQuest \cite{Simuquest:Online} Enginuity Matlab Simulink engine model with the on-line monitoring block.}
\label{Fig:exmp:simuquestEnginuityEngine}
\vspace{-12pt} 
\end{figure}

A specification of practical interest for an engine is the settling time for the A/F ratio, which is the quotient between the air mass and fuel mass flow. Ideally, the normalized A/F ratio \responsehighlight{$\lambda$} should always be 1, indicating that the ratio of the air and fuel flow is the same as the stoichiometric ratio. Under engine operating conditions, this output fluctuates $\pm \%10$. We add the on-line monitoring block to the Simulink model as presented in Fig. \ref{Fig:exmp:simuquestEnginuityEngine}.

Our goal is to monitor the engine while allowing temporary fluctuations to \responsehighlight{$\lambda$}. 
We formally define the specification as follows:

{\begin{center}
\vspace{-0.2cm}
$\phi_{pt} =  (\lambda \text{ out of bounds})  \rightarrow \Diamonddot_{[0,1]}\boxdot_{[0,1]} \neg(\lambda\text{ out of bounds}) $
\vspace{-0.2cm}
\end{center}}

Here, the formal specification states that if the A/F ratio exceeds the allowed bounds, then the ratio should have been settled for at least one second within the last two seconds.

Notice that an alternative presentation of the formula would be to use the future eventually and always operators, i.e. the formula would be defined as follows:

{\begin{center}
\vspace{-0.2cm}
$\phi_{ft} =  (\lambda \text{ out of bounds})  \rightarrow \Diamond_{[0,1]}\Box_{[0,1]} \neg(\lambda\text{ out of bounds}) $
\vspace{-0.2cm}
\end{center}}

\begin{figure}[t]
\begin{center}
\includegraphics[width= \columnwidth ]{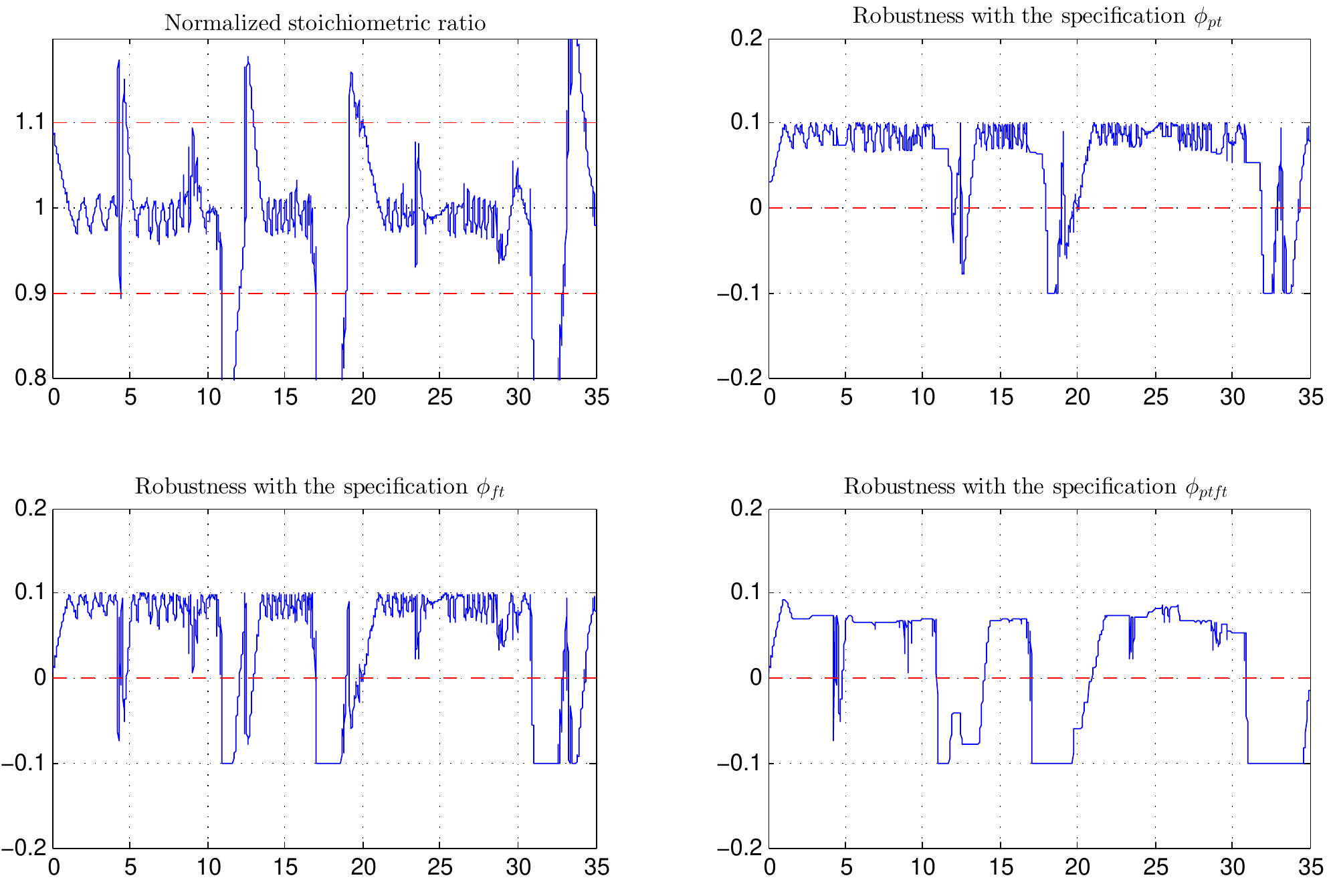}
\end{center}
\vspace{-0.5cm}
\caption{Runtime monitoring of specifications $\phi_{pt}$, $\phi_{ft}$ and $\phi_{ptft}$ on the high-fidelity engine model. The figure presents a normalized stoichiometric ratio, and the corresponding robustness values for specifications $\phi_{pt}$, $\phi_{ft}$ and $\phi_{ptft}$. Note that no predictor is utilized when computing the robustness values.}
\label{fig:simuquestMonitoring}
\vspace{-12pt} 
\end{figure}

In this case, the specification states that always, if the A/F ratio output exceeds the allowed bounds, then within one second it should settle inside the bounds and stay there for a second.

\responsehighlight{Clearly, both $\phi_{pt}$ and $\phi_{ft}$ are equivalent in terms of the set of traces that satisfy/falsify each specification}%
\footnote{Formally, this is the case if we ignore the first 2 seconds of the execution trace as well as the last 2 seconds -- if the execution trace is finite.}. \responsehighlight{However, in real-time robustness monitoring, there is an important distinction between the two. When the specification requires future information, either a predictor is put in place or the semantics will handle only the current information. In this case, without a predictor, the future time formula reduces to the propositional formula $\phi_{ft} = (\lambda \text{ out of bounds}) \rightarrow \neg (\lambda \text{ out of bounds}) \equiv (\lambda \text{ out of bounds})$. Therefore, past time operators should be used. Recall that when monitoring robustness, our goal is to provide early warning on when the specification may fail by approaching dangerously an undesired threshold. In other words, the past formula allows us to reason about the robustness of the actual system observations, while the future formula in collaboration with a forecast model would allow us to estimate the likely robustness.}
This is in contrast to many boolean monitoring algorithms which issue an ``{\it undecided until further notice}" verdict that does not provide any actionable information.

A third alternative monitoring specification is the following formula:

{\begin{center}
\vspace{-0.2cm}
$\phi_{ptft} =  \boxdot_{[0,2]}((\lambda \text{ out of bounds})  \rightarrow \Diamond_{[0,1]}\Box_{[0,1]} \neg(\lambda\text{ out of bounds})) $
\vspace{-0.2cm}
\end{center}}

This specification states that at some point in the last two seconds, when \responsehighlight{$\lambda$} is out of bounds then within the next second, \responsehighlight{$\lambda$} will not be out of bounds and stay there for one second. This alternative seems to be the balance between the \responsehighlight{$\phi_{pt}$ and $\phi_{ft}$} formulas. Where $\phi_{pt}$ purely relies on past information, and \responsehighlight{$\phi_{ft}$} relies on information from a predictor, \responsehighlight{$\phi_{ptft}$} has the advantage that it utilizes both the information from the past but also it could include information from the predictor. 

\responsehighlight{An example of real-time monitoring on the high-fidelity engine model is presented in Fig. \ref{fig:simuquestMonitoring}. 
The figure illustrates the significance of using past time operators when defining specifications. Due to the lack of predictor information, the $\phi_{ft}$ monitor falsely returns falsification at about 4 seconds whereas the $\phi_{pt}$ monitor does not.}

\responsehighlight{In the following, we analyze the overhead of the monitoring algorithm for this case study. Since the runtime is influenced by numerous sources of nondeterminism, we apply the central limit theorem to form confidence intervals for the mean simulation runtime when running the simulations with and without the monitor. To generate the results in Table \ref{tab:simuquestMonitorOverhead}, we collected 30 samples with 100 simulation runtimes in each sample. We note that the difference between the estimated mean simulation runtime when adding the monitor is 0.97\%.} The experimental results were generated on an Intel Xeon X5647 (2.993GHz, 8 CPUs) machine with 12 GB RAM, Windows 7, and Matlab 2012a.

\begin{table}[htbp]
	\vspace{-12pt}
	\setlength{\tabcolsep}{5pt}
	\renewcommand{\arraystretch}{1.5}
  \centering
  \caption{Simulation runtime statistics for the high-fidelity engine model running for 35 seconds with simulation stepsize of 0.01s. The results include the confidence intervals for the mean simulation runtime.}
  \vspace{-5pt}
    \begin{tabular}{|l|c|c|cc|cc|}
    \hline
          \multicolumn{1}{|l|}{\multirow{2}[0]{*}{Simulation runtime(sec.)}} & \multicolumn{1}{c|}{\multirow{2}[0]{*}{Est. Mean}} & \multicolumn{1}{c|}{\multirow{2}[0]{*}{Est. Std. Dev}} & \multicolumn{2}{c|}{95\%} & \multicolumn{2}{c|}{99\%} \\
          \cline{4-7}
          & \multicolumn{1}{c|}{} & \multicolumn{1}{c|}{} & LB    & UB    & LB    & UB \\
          \hline
    Without monitor & 10.811 & 0.090 & 10.778 & 10.844 & 10.766 & 10.857 \\
    With monitor & 10.987 & 0.086 & 10.955 & 11.019 & 10.944 & 11.030 \\
     \hline
    \end{tabular}%
  \label{tab:simuquestMonitorOverhead}%
  \vspace{-22pt}
\end{table}%

%% file: conclusions.tex
\section{Conclusions and Future Work}

We have presented an algorithm for monitoring the robustness of combined past and future MTL specifications.
Our framework can incorporate predicted or estimated data as provided by a model predictive component. \responsehighlight{ We have created a Simulink toolbox for MTL robustness monitoring which is distributed with the S-Taliro tools \cite{AnnapureddyLFS11tacas}.}
Our experiments indicate that the toolbox adds minimal overhead to the simulation time of Simulink models and it can be used for both runtime analysis of the models and for off-line testing.
Our future work will concentrate on several aspects. 
First, the current version of the tool allows reasoning over timed state sequences generated under a constant sampling rate.
We would like to relax this constraint so that we allow arbitrary sampling functions.
Second, we would like to investigate the possibility of porting our monitor on FPGA platforms similar to \cite{FinkbeinerK09,ReinbacherRS14tacas}.
Finally, we envision that utilizing information about the system through the form of a model will permit us to move to an event based monitoring framework while still sufficiently approximating the robustness estimate.

%% file: appendix.tex
\section*{Appendix: Proof of \yhl{Section 4.2}}

We prove by induction the correctness of Algorithms \ref{alg:on-line} and \ref{alg:comprob}. We need to prove that at each simulation step $i$, the returning value of the $CR$ algorithm is the same as the robustness value. \yhl{Without loss of generality, assume $i\ge Hst$; therefore, the values in the table columns $-Hst$ to $0$ contain the robustness values based on the actual simulation. When $i<Hst$ then the proof is immediate by the semantics of temporal logic.} We must show that, for each sub-formula $\varphi_k$ the value stored in column $j$ of robustness table $T_{k,j}$ should be correctly computed according to the semantics
$$\llbracket \varphi_k\rrbracket(s,i+j)=T_{k,j}=CR(\varphi_k,j,\responsehighlight{s'_i},\mathbf{d},\mathcal{O})$$
given matrix $T$ and vector $Pre$\\ 

{\bf Base case:} \\We will show that for each \bMITL sub-formula in the form of a predicate, the value which is returned by the CR algorithm (Algorithm \ref{alg:comprob}) is equal to the semantics of the sub-formula. Assume the sub-formula is a predicate $p=\varphi_k$, for each simulation time $i+j$, the corresponding robustness value is stored in the column $j$ of robustness table as follows: 
  $$\forall j, -Hst\le j\le Hrz, \llbracket p\rrbracket(s,i+j)=\mathbf{Dist}_d(\responsehighlight{s(i+j)},\mathcal{O}(p))=\mathbf{Dist}_d(\responsehighlight{s_{i+j}},\mathcal{O}(p))=$$$$T_{k,j}= 
  CR(p,j,\responsehighlight{s'_i},\mathbf{d},\mathcal{O})$$ 
Therefore, for each predicate the algorithm ``CR" computes the correct robustness value.\\

{\bf Induction Hypothesis:}
\\\yhl{For each temporal sub-formulas $\varphi_k$, $Hst-hst(\varphi_k)\ge Hrz$ because of the fact that \\$Hst=Hrz+hst(\varphi)\ge Hrz+hst(\varphi_k)$; therefore $-Hst+hst(\varphi_k)\le -Hrz$. \\As a result, the values at the columns from $-Hst$ up to $-Hst+hst(\varphi_k)$ will only depend on the actual simulation values, i.e., the predicates from column $-Hst$ up to column 0 which will not change in next simulation steps. These values are shown in gray color cells of Table }\ref{tab:HstHrz}. \yhl{As a result, all the table entries from $-Hst$ up to $-Hst+hst(\varphi_k)$ will not change (in next run) and the re-computation is not needed. Therefore, we shift all the values of predicates one column to the left and we ignore columns $-Hst$ to $ -Hst+hst(\varphi_k)-1$ in our current run of Algorithm}\ref{alg:on-line} \yhl{(Lines 13 and 17). Therefore, it is not necessary to include the columns $-Hst$ to $-Hst+hst(\varphi_k)-1$ in proof and Induction Hypothesis. \\For Induction Hypothesis, we assume that the value stored in the robustness table is the semantically correct robustness value for each sub-formula $\varphi_k$:}
$$\forall j, -Hst+hst(\varphi_k)\le j\le Hrz, \llbracket \varphi_k\rrbracket(s,i+j)=T_{k,j}=CR(\varphi_k,j,\responsehighlight{s'_i},\mathbf{d},\mathcal{O})$$ 
And if there exists unbounded past operator sub-formula like $\varphi_k=\psi \mathcal{S}_{[l',+\infty)} \varphi$, we assume the\yhl{ $Pre(k)=\llbracket\psi\Sc_{[l',+\infty)}\varphi\rrbracket(s,i-1-Hst+hst(\varphi_k)))$ because it belongs to the previous run of $i-1$, i.e we store the value $T_{k,(-Hst+hst(\varphi_k))}$ in $Pre(k)$ before processing the current run ($i$) (see Algorithm} \ref{alg:on-line} line 2). 

{\bf Induction Step:}
\begin{itemize}
\item [$\bullet$] Negation:
\\ $\forall j, -Hst+hst(\varphi_k)\le j\le Hrz:\llbracket\varphi_k\rrbracket(s,i+j)=\llbracket\neg\varphi_m\rrbracket(s,i+j)=$\\$-\llbracket\varphi_m\rrbracket(s,i+j)=-T_{m,j}=CR(\neg\varphi_m,j,\responsehighlight{s'_i},\mathbf{d},\mathcal{O})$
\item [] \item [$\bullet$] Disjunction:
\\$\forall j, -Hst+hst(\varphi_k)\le j\le Hrz:\llbracket\varphi_k\rrbracket(s,i+j)=\llbracket\varphi_m\vee\varphi_n\rrbracket(s,i+j)=$\\$\llbracket\varphi_m\rrbracket(s,i+j)\sqcup\llbracket\varphi_n\rrbracket(s,i+j)=T_{m,j}\sqcup T_{n,j}=CR(\varphi_m\vee\varphi_n,j,\responsehighlight{s'_i},\mathbf{d},\mathcal{O})$
\begin{table}[b]
  \centering
    \vspace{-12pt}
  \caption{Robustness Table (Unchangeable values in next runs are in gray color)}
  \fontsize{8.2}{10}
  \begin{tabular}{ c|c|c|c|c|c|c|c|c||c|c|c }
  \hline
  $column(j)\Rightarrow$&$-Hst$& ... & $-Hst+hst(\varphi_k)$ & ... & $-Hrz=-Hst+hst(\varphi)$ & ... & $-1$ & $0$ & $1$ &... & $Hrz$\\ \hline
  $index(time)\Rightarrow$&$i-Hst$& ...& $i-Hst+hst(\varphi_k)$ & ... & $i-Hrz$ & ... & $i-1$ & $i$ & $i+1$& ... & $i+Hrz$ \\ \hline
    Pre[1] &\cellcolor{gray!50}&\cellcolor{gray!50}&\cellcolor{gray!50}&\cellcolor{gray!50}&\cellcolor{gray!50}&&&&& \\ \hline
Pre[...]&&&&&&&&&& \\ \hline
 Pre[k] &\cellcolor{gray!50}&\cellcolor{gray!50}&\cellcolor{gray!50}&&&&&&& \\ \hline
  Pre[...]&&&&&&&&&& \\ \hline
  Pre[$|\varphi|$](predicate)  &\cellcolor{gray!50}&\cellcolor{gray!50}&\cellcolor{gray!50}&\cellcolor{gray!50}&\cellcolor{gray!50}&\cellcolor{gray!50}&\cellcolor{gray!50}&\cellcolor{gray!50}&&\\ \hline
  \end{tabular}
  \label{tab:HstHrz}%
  
\end{table}%

\item [] \item [$\bullet$] Until:
\\\yhl{For sub-formulas of the form $\varphi_k=\varphi_m\mathcal{U}_{[l,u]}\varphi_n$, either the corresponding robustness values are correctly saved in robustness matrix  for $\varphi_m,\varphi_n$ or the semantics will satisfy the correctness if the corresponding values belong to columns beyond the $Hrz$:} \\$\forall j, -Hst+hst(\varphi_k)\le j\le Hrz:\llbracket\varphi_k\rrbracket=\llbracket\varphi_m\mathcal{U}_{[l,u]}\varphi_n\rrbracket(s,i+j)=$\\$\sideset{}{_{h=i+j+l}^{i+j+u}}\bigsqcup(\llbracket\varphi_n\rrbracket(s,h)\sqcap\sideset{}{_{r=i+j}^{h-1}}\bigsqcap\llbracket\varphi_m\rrbracket(s,r))=$\\
$\sideset{}{_{h\in [j+l,j+u]\cap[-Hst,Hrz]}^{}}\bigsqcup(T_{n,h}\sqcap\sideset{}{_{r=j}^{h-1}}\bigsqcap T_{m,r})=CR(\varphi_m\mathcal{U}_{[l,u]}\varphi_n,j,\responsehighlight{s'_i},\mathbf{d},\mathcal{O})$


\item [] \item [$\bullet$] Bounded Since:
\\\yhl{For bounded sub-formula $\varphi_k=\varphi_m\mathcal{S}_{[l,u]}\varphi_n$, the robustness is defined as follows:
\\$\llbracket\varphi_k\rrbracket(s,i+j)=\llbracket\varphi_m\mathcal{S}_{[l,u]}\varphi_n\rrbracket(s,i+j)=\sideset{}{_{h=i+j-u}^{i+j-l}}\bigsqcup(\llbracket\varphi_n\rrbracket(s,h)\sqcap\sideset{}{_{r=h+1}^{i+j}}\bigsqcap\llbracket\varphi_m\rrbracket(s,r))$\\
Based on IH we know that $j\ge -Hst+hst(\varphi_k)$. We must show that the values of $T_{n,p}$ for $j-u\le p\le j-l$ satisfy $T_{n,p}=\llbracket\varphi_n\rrbracket(s,i+p)$ i.e. $-Hst+hst(\varphi_n)\le j-u$ and also we need to show that the values of
$T_{m,q}$ for $j-u+1\le q\le j$ satisfy $T_{m,q}=\llbracket\varphi_m\rrbracket(s,i+q)$ 
i.e. $-Hst+hst(\varphi_m)\le j-u+1$. \\\\We have two cases for $hst(\varphi_k)$:}\\
\\{\bf Case 1:} \yhl{ $hst(\varphi_k)=hst(\varphi_n)+u=max\{hst(\varphi_n)+u,hst(\varphi_m)+u-1\}$\\
According to IH, $j\ge -Hst+hst(\varphi_k)$, then $j\ge -Hst+hst(\varphi_n)+u$. Thus $j-u\ge -Hst+hst(\varphi_n)$ which satisfies the fact that $T_{n,p}=\llbracket\varphi_n\rrbracket(s,i+p)$ for $j-u\le p\le j-l$. On the other hand, in this case: $hst(\varphi_n)+u\ge hst(\varphi_m)+u-1$\\According to IH, $j+Hst\ge hst(\varphi_k)\ge hst(\varphi_m)+u-1$, i.e., $j+Hst\ge hst(\varphi_m)+u-1$. Thus $j-u+1\ge-Hst +hst(\varphi_m)$, which satisfies the fact that  $T_{m,q}=\llbracket\varphi_m\rrbracket(s,i+q)$ for $j-u+1\le q\le j$.}
\\\\{\bf Case 2:} \yhl{ $hst(\varphi_k)=hst(\varphi_m)+u-1=max\{hst(\varphi_n)+u,hst(\varphi_m)+u-1\}$\\
According to IH, $j\ge -Hst+hst(\varphi_k)$ then $j\ge -Hst+hst(\varphi_m)+u-1$. Thus $j-u+1\ge-Hst +hst(\varphi_m)$ which satisfies the fact that  $T_{m,q}=\llbracket\varphi_m\rrbracket(s,i+q)$ for $j-u+1\le q\le j$. On the other hand, in this case: $hst(\varphi_m)+u-1\ge hst(\varphi_n)+u$.\\According to IH,
$j+Hst\ge hst(\varphi_k)\ge hst(\varphi_n)+u$ i.e $j+Hst\ge hst(\varphi_n)+u$. Thus $j-u\ge -Hst+hst(\varphi_n)$ which satisfies the fact that $T_{n,p}=\llbracket\varphi_n\rrbracket(s,i+p)$ for $j-u\le p\le j-l$.
\\\\ As a result:}
  \\$\forall j, -Hst+hst(\varphi_k)\le j\le Hrz:\llbracket\varphi_k\rrbracket(s,i+j)=\llbracket\varphi_m\mathcal{S}_{[l,u]}\varphi_n\rrbracket(s,i+j)=$\\$\sideset{}{_{h=i+j-u}^{i+j-l}}\bigsqcup(\llbracket\varphi_n\rrbracket(s,h)\sqcap\sideset{}{_{r=h+1}^{i+j}}\bigsqcap\llbracket\varphi_m\rrbracket(s,r))=$\\
$\sideset{}{_{h=j-u}^{j-l}}\bigsqcup(T_{n,h}\sqcap\sideset{}{_{r=h+1}^{j}}\bigsqcap T_{m,r})=CR(\varphi_m\mathcal{S}_{[l,u]}\varphi_n,j,\responsehighlight{s'_i},\mathbf{d},\mathcal{O})$
\item [] \item [$\bullet$] Unbounded Since:
\\\yhl{For unbounded sub-formula $\varphi_k=\varphi_m\mathcal{S}_{[l,+\infty)}\varphi_n$, according to Induction Hypothesis: $$Pre(k)=\llbracket\varphi_m\Sc_{[l,+\infty)}\varphi_n\rrbracket(s,i-1-Hst+hst(\varphi_k))$$In dynamic programming we recursively update the value\\ $\llbracket\varphi_m\Sc_{[l,+\infty)}\varphi_n\rrbracket(s,i-Hst+hst(\varphi_k)+x)$\\ given the previous robustness value in the table\\ $\llbracket\varphi_m\Sc_{[l,+\infty)}\varphi_n\rrbracket(s,i-Hst+hst(\varphi_k)+x-1)$\\ (where $x=0$ when we use the $Pre(k)$)\\
\\
According to Def. 3 the robustness semantics at time $i+j$:\\
$\llbracket\varphi_m\mathcal{S}_{[l,+\infty)}\varphi_n\rrbracket(s,i+j)=\sideset{}{_{h=0}^{i+j-l}}\bigsqcup\bigg(\llbracket\varphi_n\rrbracket(s,h)\sqcap\sideset{}{_{r=h+1}^{i+j}}\bigsqcap\llbracket\varphi_m\rrbracket(s,r)\bigg)$
\\and robustness for previous time $i+j-1$:\\
$\llbracket\varphi_m\mathcal{S}_{[l,+\infty)}\varphi_n\rrbracket(s,i+j-1)=\sideset{}{_{h=0}^{i+j-l-1}}\bigsqcup\bigg(\llbracket\varphi_n\rrbracket(s,h)\sqcap\sideset{}{_{r=h+1}^{i+j-1}}\bigsqcap\llbracket\varphi_m\rrbracket(s,r)\bigg)$
\\We can define robustness value at time $i+j$ given the value at time $i+j-1$:\\

$\llbracket\varphi_m\mathcal{S}_{[l,+\infty)}\varphi_n\rrbracket(s,i+j)=\sideset{}{_{h=0}^{i+j-l}}\bigsqcup\bigg(\llbracket\varphi_n\rrbracket(s,h)\sqcap\sideset{}{_{r=h+1}^{i+j}}\bigsqcap\llbracket\varphi_m\rrbracket(s,r)\bigg)=$\\\\
$=\bigg(\sideset{}{_{h=0}^{i+j-l-1}}\bigsqcup\Big(\llbracket\varphi_n\rrbracket(s,h)\sqcap\sideset{}{_{r=h+1}^{i+j-1}}\bigsqcap\llbracket\varphi_m\rrbracket(s,r)\Big)\sqcap\llbracket\varphi_m\rrbracket(s,i+j)\bigg)\bigsqcup$\\$\Big(\llbracket\varphi_n\rrbracket(s,i+j-l)\sqcap\sideset{}{_{r=i+j-l+1}^{i+j}}\bigsqcap\llbracket\varphi_m\rrbracket(s,r)\Big)=$\\\\
$=\bigg(\llbracket\varphi_m\mathcal{S}_{[l,+\infty)}\varphi_n\rrbracket(s,i+j-1)\sqcap\llbracket\varphi_m\rrbracket(s,i+j)\bigg)\bigsqcup$\\$\Big(\llbracket\varphi_n\rrbracket(s,i+j-l)\sqcap\sideset{}{_{r=i+j-l+1}^{i+j}}\bigsqcap\llbracket\varphi_m\rrbracket(s,r)\Big)$\\\\
Based on IH, we know that $j\ge -Hst+hst(\varphi_k)$. We must show that the value of $T_{n,j-l}=\llbracket\varphi_n\rrbracket(s,i+j-l)$, i.e., $-Hst+hst(\varphi_n)\le j-l$ and also the values of $T_{m,q}$ for $j-l+1\le q\le j$ satisfy $T_{m,q}=\llbracket\varphi_m\rrbracket(s,i+q)$, i.e., $-Hst+hst(\varphi_m)\le j-l+1$.
\\\\We have two cases for $hst(\varphi_k)$:}\\
\\{\bf Case 1:} \yhl{ $hst(\varphi_k)=hst(\varphi_n)+l=max\{hst(\varphi_n)+l,hst(\varphi_m)+l-1\}$\\
According to IH, $j\ge -Hst+hst(\varphi_k)$; therefore, $j\ge -Hst+hst(\varphi_n)+l$ and $j-l\ge -Hst+hst(\varphi_n)$ which satisfies $T_{n,j-l}=\llbracket\varphi_n\rrbracket(s,i+j-l)$. On the other hand in this case: $hst(\varphi_n)+l\ge hst(\varphi_m)+l-1$.
\\According to IH, $j+Hst\ge hst(\varphi_k)\ge hst(\varphi_m)+l-1$, i.e., $j+Hst\ge hst(\varphi_m)+l-1$. Thus $j-l+1\ge -Hst+hst(\varphi_m)$ which satisfies the fact that  $T_{m,q}=\llbracket\varphi_m\rrbracket(s,i+q)$ for $j-l+1\le q\le j$.
}
\\\\{\bf Case 2:} \yhl{ $hst(\varphi_k)=hst(\varphi_m)+l-1=max\{hst(\varphi_n)+l,hst(\varphi_m)+l-1\}$\\
According to IH, $j\ge -Hst+hst(\varphi_k)$; therefore, $j\ge -Hst+hst(\varphi_m)+l-1$ where $j-l+1\ge -Hst+hst(\varphi_m)$ which satisfies the fact that  $T_{m,q}=\llbracket\varphi_m\rrbracket(s,i+q)$ for $j-l+1\le q\le j$. On the other hand in this case: $hst(\varphi_m)+l-1\ge hst(\varphi_n)+l$.
\\According to IH, $j+Hst\ge hst(\varphi_k)\ge hst(\varphi_n)+l$ i.e. $j+Hst\ge hst(\varphi_n)+l$.\\
thus $j-l\ge -Hst+hst(\varphi_n)$ which satisfies $T_{n,j-l}=\llbracket\varphi_n\rrbracket(s,i+j-l)$\\\\
 As a result:}
\\$\forall j, -Hst+hst(\varphi_k)\le j\le Hrz:\llbracket\varphi_k\rrbracket(s,i+j)=\llbracket\varphi_m\mathcal{S}_{[l,+\infty)}\varphi_n\rrbracket(s,i+j)=$\\$=\bigg(\llbracket\varphi_m\mathcal{S}_{[l,+\infty)}\varphi_n\rrbracket(s,i+j-1)\sqcap\llbracket\varphi_m\rrbracket(s,i+j)\bigg)\bigsqcup$\\$\Big(\llbracket\varphi_n\rrbracket(s,i+j-l)\sqcap\sideset{}{_{r=i+j-l+1}^{i+j}}\bigsqcap\llbracket\varphi_m\rrbracket(s,r)\Big)=$\\

$=\bigg(\left\{ \begin{array}{ll}
T_{k,j-1} & \mbox{ if }  j>-Hst+hst(\varphi_k) \\
 Pre[k] & \mbox{ if }  j=-Hst+hst(\varphi_k) \\
\end{array}\right\}\sqcap T_{m,j}\bigg)\bigsqcup\Big(T_{n,j-l}\sqcap\sideset{}{_{r=j-l+1}^{j}}\bigsqcap T_{m,r}\Big)=\\
\Big(tmp_{\Sc}\Big)\bigsqcup\Big(T_{n,j-l}\sqcap tmp_{min}\Big) =CR(\varphi_m\mathcal{S}_{[l,+\infty)}\varphi_n,j,\responsehighlight{s'_i},\mathbf{d},\mathcal{O})$.

\end{itemize}